\documentclass[twocolumn,showpacs,amsmath]{revtex4}
\usepackage{graphicx}%
\usepackage{dcolumn}
\usepackage{amsmath}
\usepackage{here}

\begin{document}

\title{Waveguiding properties of optical vortex solitons}

\author{Jos\'e R. Salgueiro and Yuri S. Kivshar}

\affiliation{Nonlinear Physics Group, Research School of Physical
Sciences and Engineering, Australian National University, Canberra
ACT 0200, Australia}

\begin{abstract}
We study the properties of linear and nonlinear waveguides induced
by {\em optical vortex solitons} in both self-defocusing and
self-focusing nonlinear media, for the case of a saturable
nonlinear response. We demonstrate that the vortex-induced
waveguides can guide both fundamental and first-order guided modes
which together with the vortex may form, for larger amplitudes of
the guide modes, different types of composite vortex-mode vector
solitons. In the case of the self-focusing saturable medium, we
demonstrate that a large-amplitude guided mode can stabilize the
ring-like vortex structure which usually decays due to azimuthal
modulational instability.
\end{abstract}

\maketitle

\section{Introduction}

Vortices are fundamental localized objects which appear in many
branches of physics, ranging from the physics of liquid crystals
to the dynamics of superfluids and Bose-Einstein
condensates~\cite{pismen}. Different types of vortices can be
found and identified in optics; one of the simplest objects of
this kind is {\em a phase singularity} in an optical wave front
which is associated with a phase dislocation carried by a
diffracting optical beam~\cite{review}.

When such an optical vortex propagates in {\em self-defocusing
nonlinear medium}, the vortex core with a phase dislocation
becomes self-trapped, and the resulting {\em stationary singular
beam} is known as {\em an optical vortex soliton}~\cite{swartz}.
Such optical vortex solitons exist on a nonvanishing background
wave, and they represent a two-dimensional generalization of the
so-called {\em dark solitons}~\cite{book}. They can be generated
in experiment as self-trapped singularities of broad beams by
using {\em a holographic phase mask}, as was demonstrated for
different types of nonlinear defocusing
media~\cite{exp1,exp2,exp3}. As a matter of fact, this type of
optical vortex solitons demonstrates many common features with the
vortices observed in superfluids and Bose-Einstein
condensates~\cite{bec}.

In {\em self-focusing nonlinear media}, optical vortices can exist
as ring-like optical beams with zero intensity at the center
carrying a phase singularity~\cite{kruglov}. However, due to the
self-focusing nature of nonlinearity such ring-like vortex beams
become unstable, and they are known to decay into several
fundamental optical solitons flying off the main
ring~\cite{firth}. This effect was observed experimentally in
different nonlinear systems, including the saturable Kerr-like
nonlinear media, biased photorefractive crystals, and quadratic
nonlinear media in the self-focusing regime (see details and
references in Ref.~\cite{book}).

Waveguides induced by optical vortices in both linear and
nonlinear regimes are of a special interest because this type of
waveguides is robust and can be made reconfigurable. Earlier
numerical and experimental results~\cite{wave1,wave2,wave3}
indicate that many of the vortex waveguiding properties should be
similar to those of planar dark solitons~\cite{book}. Moreover,
the vortex-induced waveguides can guide large-amplitude beams
beyond the applicability limits of the linear guided-wave theory,
and, together with the guided beam, they can form a special type
of {\em the vortex-mode vector soliton} or its various
generalizations~\cite{vortex,dipole,dipole2,dipole3}.

In this paper, we carry out a systematic analysis of the
waveguiding properties of the vortex solitons and vortex-mode
vector solitons in saturable nonlinear media, for both
self-defocusing and self-focusing nonlinear media. We follow the
earlier analysis of Ref.~\cite{wave3} and examine {\em two major
regimes} of the vortex waveguiding. In the linear regime, the
guided beam is weak, and the analysis of the vortex waveguiding
properties is possible by approximate analytical methods, reducing
the problem to a well--known analysis of the linear guided-wave
theory~(as an example, see Ref.~\cite{wave2}). The most
interesting nonlinear regime corresponds to large intensities of
the guided beam,  and it gives rise to composite (or vector)
solitons with a vortex component, that should be identified and
analyzed numerically.

We also analyze the vortex waveguiding and vortex-mode vector
solitons in a self-focusing nonlinear medium where, as is well
known, the vortex beam becomes self-trapped and it also displays
strong azimuthal modulational instability. However, we demonstrate
that a mutual incoherent coupling between the vortex waveguide and
a large-amplitude guided mode it guides can provide a strong
stabilizing mechanism for stable two-component vortex solitons to
exist in such media.

The paper is organized as follows. The next section (Sec. II)
presents our model which takes into account both the incoherent
coupling between the waveguide and the localized mode it guides,
and the saturable nature of nonlinearity usually realized in
experiment. The model is valid for both self-defocusing and
self-focusing media. Section III is devoted to the study of the
properties of optical vortex solitons and their guiding properties
in self-defocusing media, whereas Sec. IV summarizes our results
for the vortex waveguides and vector solitons in self-focusing
nonlinear media. In particular, for the first time to our
knowledge, we demonstrate that the vortex-mode vector solitons
formed by the vortex waveguide together with a large-amplitude
mode it guides can show the {\em bistability} property, when two
types of different composite soliton structures exist for the same
value of the soliton power. We also discuss, by means of direct
numerical simulations, the stability properties of the vortex-mode
vector solitons in a self-focusing nonlinear medium. In
particular, we reveal an effective method of suppression of the
azimuthal instability through the stabilization effect that the
guided mode exerts on the hosting vortex, the latter is known to
be unstable by itself in focusing media. Finally, Sec. V concludes
our paper.

\section{Model}

In  order to study the guided modes of a vortex-induced waveguide
in both self-focusing and self-defocusing media, we consider the
interaction of two mutually incoherent optical beams propagating
in a nonlinear saturable medium. The equations of motion for two
beams can be presented in the dimensionless form as follows,
\begin{equation}
\label{vortex_induced_vaweguides}
   \begin{array}{l} {\displaystyle
      i \frac{\partial u}{\partial z} +\Delta_{\perp} u + \eta \,
        \frac{(|u|^2 + \mu |v|^2) u}{1+\sigma (|u|^2+|v|^2)} = 0,
    } \\*[9pt] {\displaystyle
      i \frac{\partial v}{\partial z} +\Delta_{\perp} v + \eta \,
        \frac{(|v|^2 + \mu |u|^2) v}{1+\sigma (|u|^2+|v|^2)} = 0,
      }
   \end{array}
\end{equation}
where $u$ and $v$ are the dimensionless amplitudes of two fields,
the parameter $\sigma$ characterizes the nonlinearity saturation
effect, and the incoherent mode interaction is described by the
coupling parameter $\mu$. The spatial coordinate $z$  is the
propagation direction of the beams, and $\Delta_\perp$ stands for
the transversal part of the Laplacian operator in the cylindrical
coordinates $r=(x^2+y^2)^{1/2}$ and $\phi =\tan^{-1}(y/x)$. The
sign parameter $\eta=\pm 1$ defines the type of the nonlinear
medium under consideration, which is self-focusing, for $\eta=+1$,
or self-defocusing, for $\eta = -1$, respectively.

The model (\ref{vortex_induced_vaweguides}) provides a
straightforward generalization to a number of important cases
studied earlier. In particular, the limit $\sigma \rightarrow 0$
corresponds to the Kerr medium discussed, for example, in
Refs.~\cite{wave1,wave3,haelt}, whereas the saturable case at
$\mu=1$ corresponds to the incoherent beam interaction in
photorefractive nonlinear media where different types of composite
vector solitons have been predicted theoretically and observed in
experiment~\cite{vortex,dipole,dipole2,dipole3}. In particular,
some examples of the two-component vector solitons composed of a
vortex component and the localized modes it guides were presented
earlier by Haelterman and Sheppard~\cite{haelt} for the defocusing
Kerr nonlinearity.

We look for stationary solutions of the system
\ref{vortex_induced_vaweguides} in the form of a radially
symmetric single-charged vortex beam in the field $u$ of the form
\begin{equation}
\label{eq_u} u (r, \phi; z)=u(r)e^{i\phi}e^{i \gamma z},
\end{equation}
where $\gamma$ is the propagation constant of the vortex mode,
which if considered normalized takes the values +1 or -1 in the
focusing and defocusing cases respectively, and consequently we
have $\gamma=\eta$. For $r\rightarrow \infty$, the amplitude
$u(r)$ approaches a constant value, for the self-defocusing
medium, or zero, for the self-focusing case, respectively.  We
assume that the vortex guides the second beam given by the
expression
\begin{equation}
\label{eq_v}
 v(r, \phi; z)=v(r)e^{il\phi}e^{i\beta z},
 \end{equation}
 where $\beta$ is the propagation constant of the guided beam.
 In the expressions (\ref{eq_u}),
(\ref{eq_v}) the functions $u(r)$ and $v(r)$ are the  radial
envelopes of the interacting fields, and $l$ is the angular
momentum of the guided mode.

\begin{figure}
\centerline{\includegraphics[width=3.2in]{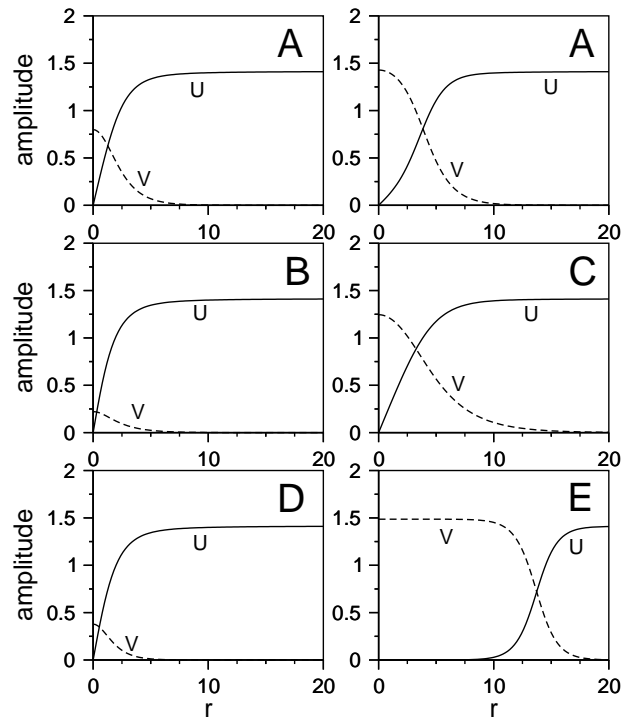}}
\caption{Examples of localized solutions for the vortex waveguide
($u$) guiding a fundamental-mode beam ($v$). The labels correspond
to the points marked in Fig.~\ref{fig1}. Two examples labelled by
the same letter A constitute two different types of localized
solutions found in this domain (mode bistability). The parameters
are:  A ($\mu =1.5$, $\beta =-1.1$), B ($\mu =1.1$, $\beta
=-0.85$), C ($\mu =1.1$, $\beta =-1.0$), D ($\mu =2.0$, $\beta
=-1.3$), and E ($\mu =2.0$, $\beta =-1.05$). } \label{fig2}
\end{figure}

In this way, the following set of stationary $z$-independent
equations are obtained:
\begin{equation}
\label{stationary_vortex_induced_vaweguides}
   \begin{array}{l} {\displaystyle
      - \eta u +\Delta_r u -\frac{1}{r^2}u + \eta \,
        \frac{(u^2 + \mu v^2) u}{1+\sigma (u^2+v^2)} = 0,
    } \\*[9pt] {\displaystyle
      -\beta v +\Delta_r v -\frac{l^2}{r^2}v + \eta \,
        \frac{(v^2 + \mu u^2) v}{1+\sigma (u^2+v^2)} = 0,
      }
   \end{array}
\end{equation}
where $\Delta_r \equiv (1/r)d/dr(r~d/dr)$. The model
\ref{stationary_vortex_induced_vaweguides} describes different
types of localized solutions for the vortex and the localized
modes it guides, and it has been further analyzed by the numerical
relaxation methods to classify its localized solutions for both
defocusing and focusing cases, respectively.

In the defocusing case when $\eta=-1$, we have $u \rightarrow u_0$
and  $v \rightarrow 0$ for $r \rightarrow \infty$. Using this
asymptotic conditions, we obtain
\begin{equation}
\left(1-\frac{u_0^2}{1+\sigma u_0^2}\right) u_0=0,
\end{equation}
and find the result for the vortex background amplitude $u_0$ in
the form, $u_0 =(1-\sigma)^{-1/2}$.

Using the second equation for $v$, and neglecting the terms which
most quickly tend to zero when $r\rightarrow \infty$, we obtain:
\begin{equation}
-\beta v + \frac{d^2v}{dr^2}-\frac{\mu u_0^2}{1+\sigma u_0^2} v
=0.
\end{equation}
In order to have the exponentially decaying function $v$ (for
$r\rightarrow \infty$), the following condition should be
satisfied,
\begin{equation}
-\beta + \frac{\mu u_0^2}{1+\sigma u_0^2} <0.
\end{equation}
that gives as the result
\begin{equation}
\beta + \mu >0 \Longrightarrow \beta > -\mu,
\end{equation}
which defines the existence domain threshold as the line
$-\beta=\mu$. This result is valid for both fundamental and
first-order guided modes.


\section{Vortex guided modes}

First, we consider the case of self-defocusing nonlinearity when
$\eta =-1$. In this case, the vortex amplitude $u(r)$ approaches a
finite-amplitude asymptotic value $u_0$ for $r \rightarrow
\infty$, where $u_0$ can be found from the asymptotic analysis,
$u_0 = (1-\sigma)^{-1/2}$. With the help of numerical relaxation
methods, we find various types of localized modes that describe a
vortex and the localized guided mode with the amplitude $v(r)$
satisfying the condition $v(r) \rightarrow 0$ for $r \rightarrow
\infty$. This condition requires that $\beta<0$, since the field
$v(r)$  is localized due to the presence of the vortex waveguide
only, and it describes a guided mode.

\begin{figure}
\centerline{\includegraphics[width=3.2in]{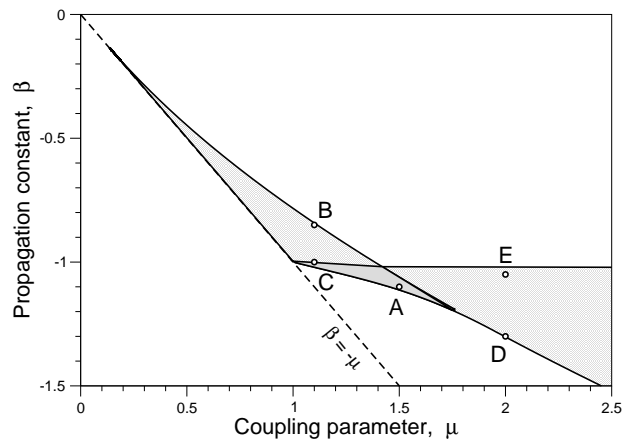}}
\caption{Existence domain of the fundamental mode in the
vortex-induced waveguides on the plane ($\mu$, $\beta$), for
$\sigma=0.5$. Marked points correspond to the solutions presented
above in Fig.~\ref{fig2}. A triangle region with the point A
corresponds to the existence domain for bistable composite
solitons.} \label{fig1}
\end{figure}

\begin{figure}
\centerline{\includegraphics[width=3.2in]{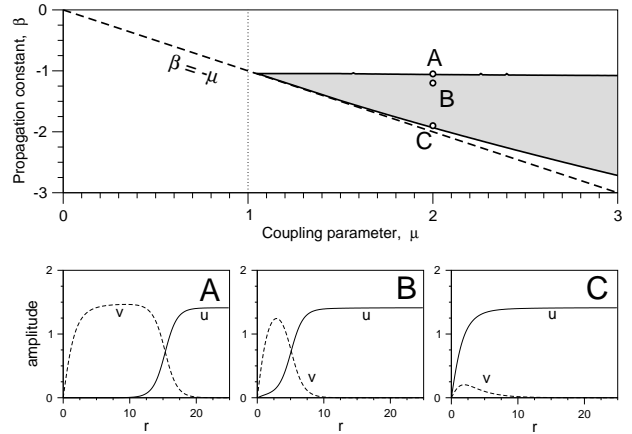}}
\caption{Top: existence domain for the first-order mode ($l=1$)
guided by the vortex-induced waveguide, show on the plane ($\mu$,
$\beta$). Below: examples of the two-component localized solutions
for $\sigma =0.5$,  $\mu =2$, and $\beta = -1.05$ (A), $\beta
=-1.2$ (B), and $\beta=-1.9$ (C), respectively.} \label{fig3}
\end{figure}

Figure~\ref{fig2} shows several characteristic solutions of this
type for particular values of $\mu$ and $\sigma=0.5$, whereas the
existence domain for the solutions with the fundamental guided
mode (i.e. when $l=0$) is shown in Fig.~\ref{fig1}. For the values
of the propagation constant $\beta$ close to the cutoff ($\beta_c
\approx -0.79$ at $\mu =1$), the second field presents a
low-amplitude guided mode that can be analyzed by means of the
linear guided wave theory. However, for values of $\beta$ close to
the threshold, the amplitude of the guided mode becomes comparable
to that of the vortex and its width grows, as shown for the case E
in Fig.~\ref{fig2}. In this latter case, the vortex is deformed as
well, so the resulting structure describes a kind of a
two-component vector soliton with the vortex-like mode.

In Fig.~\ref{fig3}, we show the existence domain for the first
vortex-like ($l=1$) mode of the induced vortex waveguide. Similar
to the case of the fundamental guided mode, the linear theory is
applicable near the mode cut-off, as is shown for the mode C,
where the mode grows and it deformes the vortex waveguide
substantially, see the case A near the upper boundary of the
existence domain in Fig.~\ref{fig3} (top).

\section{Self-trapped vortex beams}
\label{self_trapped_vortices}

\subsection{Vortex-mode vector solitons}

In the case of the self-focusing nonlinear medium, we put
$\eta=+1$,  and consider the self-trapped vortex beams with the
asymptotic behavior  $u(r) \rightarrow 0$ as $r \rightarrow
\infty$. Besides that, it is required that  $v(r) \rightarrow 0$
and  $\beta>0$, so that $v(r)$ describes a localized mode. At the
origin ($r=0$), the boundary condition for the vortex is $u=0$
and, if we seek the solutions corresponding to the fundamental
guided mode ($l=0$), we have $v=v_0$ (or $dv/dr=0$) for the field
$v(r)$ at $r=0$.

\begin{figure}
\centerline{\includegraphics[width=3.2 in]{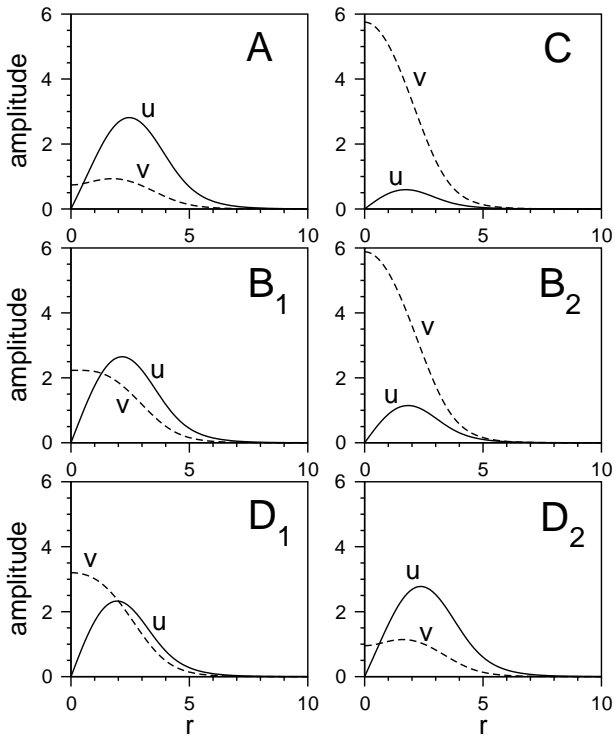}}
\caption{Examples of solutions for the fundamental mode of a
waveguide induced by a bright-like vortex. Labels correspond to
the points marked in FIG. 4, using a subscript to name the two
solutions found for the points B and C. Values: A $(\mu=1.10,
\beta=1.38)$, B $(\mu=1.10, \beta=1.44)$, C $(\mu=1.17,
\beta=1.38)$, D $(\mu=1.17, \beta=1.47)$. } \label{fig5}
\end{figure}

\begin{figure}
\centerline{\includegraphics[width=3.2 in]{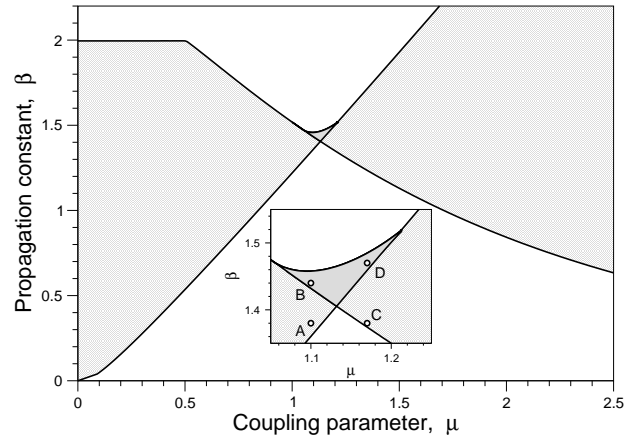}}
\caption{Existence domain in the plane ($\mu$,$\beta$) of the
fundamental mode guided by the self-trapped vortex  for
$\sigma=0.5$. Inset shows a zoomed region where two different
types of solutions exist (bistability zone). } \label{fig4}
\end{figure}

In Fig.~\ref{fig5}, we show some examples of the two-component
localized solutions which describes a fundamental (no nodes) beam
guided by the self-trapped vortex in the form of a vortex-mode
soliton. The existence domain for this kind of solutions is shown
in Fig.~\ref{fig4} on the parameter plane ($\mu$, $\beta$),
calculated numerically for $\sigma=0.5$. Close to the cutoff, the
guided mode has a small amplitude (see the cases A and D$_2$), and
the vortex is only weakly distorted. However, when the propagation
constant $\beta$ is close to the upper threshold,  the field
$v(r)$ grows in the amplitude affecting strongly the vortex mode.

\begin{figure}
\centerline{\includegraphics[width=3.2 in]{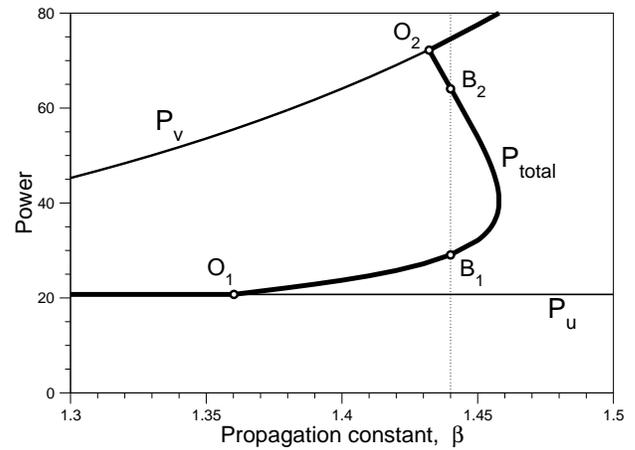}}
\caption{Bifurcation diagram for the one-and two-component
solitons composed by the bright vortex inducing a waveguide
together with its fundamental mode for $\mu=1.1$. Thin lines
represent the powers $P_u$ and $P_v$ of each of the scalar
solitons; the thick line is the total power $P_{\rm total}$ of the
composite soliton. The points O$_1$ and O$_2$ are the bifurcation
points where the vector soliton emerges, and the points B$_1$ and
B$_2$ correspond to the examples of the bistable solutions for
$\beta=1.44$ shown in Fig.~\ref{fig5}.} \label{fig6}
\end{figure}

\begin{figure}
\centerline{\includegraphics[width=3.2in]{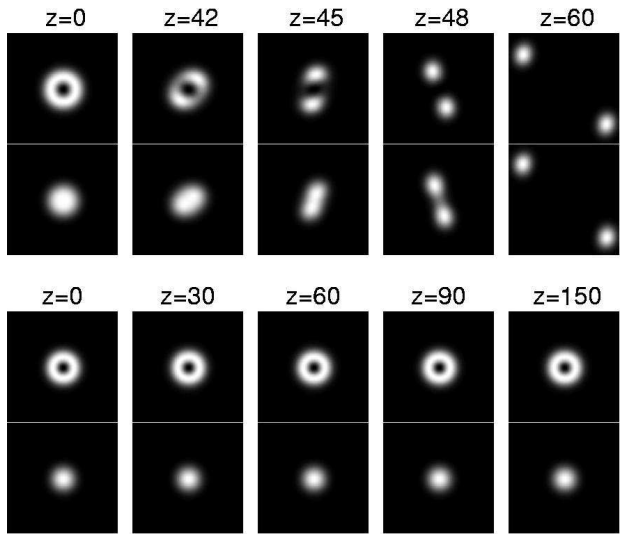}}
\caption{Examples of the vortex evolution in the bistability
domain. Shown are the field intensity profiles as gray-scale
images at several propagation distances. Top row: the components
of the vector soliton labelled as B$_1$ in Fig.~\ref{fig6}. Bottom
row: the components of the vector soliton labelled as B$_2$ in
Fig.~\ref{fig6}.} \label{fig9}
\end{figure}

In order to describe the bistable vector solitons, in
Fig.~\ref{fig6} we plot the bifurcation diagram of the
two-component localized solutions, for the partial and total beam
powers. The power of the composite vortex-mode solitons $P_{\rm
total}$ originates at the bifurcation point $O_1$ where the mode
$v$ is small and it can be described by the linear theory. For
larger value of $\beta$, the curve bends, and then it merges with
the other partial power curve $P_v$ at the bifurcation point $O_2$
(see Fig.~\ref{fig6}). The bistable solutions B$_1$ and B$_2$
correspond to a single value of the propagation constant $\beta$
in the bistability domain. Importantly, two solutions have
different stability properties, and only one of the solutions is
stable, as shown in Fig.~\ref{fig9}.

\subsection{Vortex stabilization}

The incoherent interaction between the vortex and the localized
mode it guides has the character of attraction, and it may provide
an effective physical mechanism for stabilizing the vortex beam in
a self-focusing nonlinear medium. Indeed, it is well-known that
the vortex beam becomes unstable in a self-focusing nonlinear
medium due to the effect of the azimuthal modulational
instability. In this case, the vortex splits into the fundamental
beams that fly off the main vortex ring~\cite{firth}. On the other
hand, the bright solitons are known to be stable in such media. We
expect that a mutual attraction of the components in a
two-component system would lead to a counter-balance of the vortex
instability by the bright component if the amplitude of the latter
is large enough.

To carry out this study, we consider a two-component composite
structure consisting of a vortex beam together with the
fundamental mode of the waveguide it guides, described by
Eqs.~\ref{vortex_induced_vaweguides} (at $\eta=1$ and $\mu=1.0$).
To study the mode stability, we propagate the stationary soliton
solutions numerically.

\begin{figure} \centerline{\includegraphics[width=3.2
in]{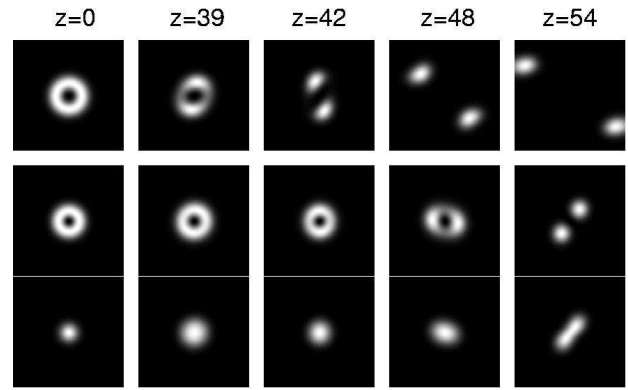}} \caption{Vortex stabilization in a
two-component system. Shown are the field intensity profiles as
gray-scale images at several propagation distances. Top row:
propagation of a vortex soliton by itself ($v=0$). Bottom row:
both the vortex beam and fundamental mode (approximated by a
Gaussian profile) propagate together. Parameters are:
$\sigma=0.5$, $\mu=1$, and $\beta=1.45$.} \label{fig7}
\end{figure}

In Fig.~\ref{fig7}, we compare the vortex stability for the scalar
and two-component systems. In the top row, we show the propagation
of a vortex alone in the scalar model; the vortex breaks into two
solitons which fly away after some propagation distance. In the
bottom row, we show the propagation of two coupled components (the
vortex and bright mode it guides). Due to a strong coupling
between the modes, the propagation of the vortex is stabilized and
its decay can be delayed dramatically, as shown in
Fig.~\ref{fig7}(lower row), or become completely stable, similar
to the case shown in Fig.~\ref{fig9}(low row). We confirm this
stabilization mechanism by performing a similar study for a
Gaussian input beam of the bright component instead of the exact
stationary state, as is easier to realize in experiment.

\section{Conclusions}

We have studied the properties of the vortex-induced optical
waveguides in both self-focusing and self-defocusing saturable
nonlinear media. We have calculated numerically the existence
domains of the composite solitons created by an optical vortex and
the localized mode it guides, and have analyzed the properties of
the vortex waveguides in both linear and nonlinear regimes. In the
case of self-defocusing nonlinear media, we have identified the
regimes when the vortex soliton can guide both the fundamental and
first-order guided modes creating an effective multi-mode
waveguide. In the case of self-focusing nonlinear media, we have
studied the effect of the  strong coupling between the vortex
waveguide and the guided mode on the vortex stability. In
particular, we have described a novel mechanism of stabilizing the
vortex azimuthal modulational instability by a large-amplitude
guide mode, and we have revealed the existence of optical
bistability for the vortex-mode composite solitons. We believe
similar results can be obtained for vortices in other types of
nonlinear models, and they can be useful for other fields such as
the physics of the Bose-Einstein condensates of ultra-cold atoms.

\section*{Acknowledgements}

We thank Sergey Odoulov and Mikhail Vasnetsov for the invitation
to submit our contribution to the special issue, and we dedicate
this article to Professor Marat Soskin, a pioneer of the physics
of optical vortices and singular optics, on the occasion of his
75-th birthday.

This work was partially supported by the Australian Research
Council. J.R. Salgueiro acknowledges a posdoctoral fellowship
granted by the Secretar\'{\i}a de Estado de Educaci\'on y
Universidades of Spain and partially supported by the European
Social Fund.

\end{document}